\documentclass{WileyChemistry-template}
\usepackage[version=3]{mhchem} 
\usepackage[subpreambles=false]{standalone}
\usepackage{import}
\usepackage{blindtext}
\usepackage{amsmath}
\usepackage[utf8]{inputenc}
\usepackage{textcomp}
\usepackage{graphicx}
\usepackage{physics}
\usepackage{tabularx}      
\usepackage{booktabs}     
\usepackage{threeparttable}

\title
  {Molecular glues stabilize water-mediated hydrogen bonds in ternary complexes}

\author{
\begin{minipage}{\textwidth}
	Apoorva Mathur\textsuperscript{a}, 
    Mariona Alegre Canela\textsuperscript{a,b},
    Max von Graevenitz\textsuperscript{a}, 
    Chiara Gerstner\textsuperscript{a},
    Ariane Nunes-Alves\textsuperscript{a,}*
\end{minipage}
}

\newcommand{\affiliation}{
\begin{itemize}

\item[{a}] A. Mathur, M. A. Canela,  M. v. Graevenitz, C. Gerstner, A. Nunes-Alves\\
Institute of Chemistry, Technische Universität Berlin, Straße des 17. Juni 135, 10623 Berlin, Germany\\
E-mail: ferreira.nunes.alves@tu-berlin.de \\

\item[{b}] M. A. Canela\\
Universitat Politècnica de Catalunya,  Building A (EEBE) Av. Eduard Maristany, 16, 08019, Barcelona, Spain \\

\end{itemize}
}

\renewcommand{\dedication}{
	\begin{minipage}{\textwidth}
	\end{minipage}
}

\renewcommand{\abstract}{
By stabilizing weak and transient protein-protein interactions (PPIs), molecular glues address the challenge of targeting proteins previously considered undruggable. Rapamycin and WDB002 are molecular glues that bind to FK506-binding protein (FKBP12) and target the FKBP12-rapamycin-associated protein (FRAP) and the centrosomal protein 250 (CEP250), respectively. Here, we used molecular dynamics simulations to gain insights into the effects of molecular glues on protein conformation and PPIs. The molecular glues modulated protein flexibility, leading to less flexibility in some regions, and changed the pattern and stability of water-mediated hydrogen bonds between the proteins. Our findings highlight the importance of considering water-mediated hydrogen bonds in developing strategies for the rational design of molecular glues. 

}

\newcommand{\keywords}{
	\textbullet\ Molecular dynamics simulations 
	\textbullet\ Molecular glue
	\textbullet\ Ternary complex
	\textbullet\ Hydrogen bonds
	\textbullet\ FK506-binding protein (FKBP12)
}
\begin{document}
\twocolumn[\vspace{-1.5cm}\maketitle\vspace{-1cm}
	\textit{\dedication}\vspace{0.4cm}]
\small{\begin{shaded}
		\noindent\abstract
	\end{shaded}
}
\begin{figure} [!b]
\begin{minipage}[t]{\columnwidth}{\rule{\columnwidth}{1pt}\footnotesize{\textsf{\affiliation}}}\end{minipage}
\end{figure}

\section{Introduction}

Many biological processes are regulated by an intricate network of protein-protein interactions (PPIs), which may be altered in disease conditions. Consequently, PPI modulation has been an important strategy in drug discovery \cite{konstantinidou2024molecular}. While early efforts primarily focused on developing inhibitors to block these interactions, the past decade has seen growing interest in approaches that stabilize PPIs \cite{nada2024new}. 

Two new classes of small molecules emerged in recent years to enable the formation of protein-protein complexes, or to stabilize existing ones. One of them are PROteolysis TArgeting Chimeras (PROTACs), which are small molecules with three components: a ligand binding to a target protein, a ligand binding to an effector protein, and a linker connecting these two ligands \cite{gao2020protac, as2024protac}. PROTACs stabilize the complex between a target protein and an effector protein, usually E3 ubiquitin ligase, leading to the formation of a ternary complex and degradation of the target protein through the ubiquitin-proteasome system \cite{gao2020protac, sun2019protacs}. The other class, and the focus of this work, are molecular glues, which are small molecules that, similar to PROTACs, promote binding of an effector protein to a target protein, leading to the formation of a ternary complex \cite{tomlinson2025three}. PROTACs and molecular glues hold the promise of targeting proteins which are considered "undruggable" due to the lack of deep binding sites to accommodate small molecule inhibitors \cite{holdgate2024screening,sun2019protacs, xie2023recent}. 

Previous work classified molecular glues in three main types: degradative, non-degradative or PPI stabilizers, and molecular glues that induce self-association \cite{konstantinidou2024molecular}.
Here, we investigate molecular glues which act as PPI stabilizers. Well studied examples of such PPI-stabilizing molecular glues are the macrolides rapamycin (sirolimus) and tacrolimus (FK506), which bind the immunophilin 12kDa FK506-binding protein (FKBP12) with high affinity (dissociation constant, K$_d$, of 0.2 \cite{banaszynski2005characterization} and 0.4 nM \cite{siekierka1989cytosolic}, respectively), and subsequently stabilize ternary complexes with distinct protein partners, such as the kinase domain of the FKBP12-rapamycin-associated protein (FRAP, also known as mechanistic target of rapamycin, mTOR) and calcineurin. Both ternary complexes lead to immunosuppression through different mechanisms \cite{choi1996structure}. More recently, Warp Drive Bio Inc discovered a natural product, WDB0002, that also binds to FKBP12 (K$_d$ of 5.2 $\pm$ 0.4 nM) \cite{shigdel2020genomic}. It acts as the molecular glue between FKBP12 and the human centrosomal protein 250 (CEP250), which is composed of a filamentous leucine zipper fold and disrupts NEK2-mediated centrosome separation \cite{shigdel2020genomic}.

The emergence of computational tools such as AlphaFold2 \cite{nunes2023alphafold2,xie2023recent} has facilitated the modeling of ternary complexes. While many computational tools exist for screening and optimization of small molecules \cite{salo2020molecular, schimunek2024community}, most of them were developed for binary complexes, and need to be adjusted and tested for ternary complexes. Initially, the identification of molecular glues depended on serendipity, but more systematic screening approaches are being developed \cite{rui2023protein}, and several computational methods were recently proposed for the design, screening, optimization and modelling of ternary complexes mediated by molecular glues \cite{ben2025molecular,xu2025accommodating,deutscher2025discovery,lukauskis2025optimizing}. Rui et al \cite{rui2023protein} collected all ternary complexes mediated by molecular glues with experimental structures available, and used this data to investigate the properties of PPI interfaces, what led to the classification of the ternary complex structures in two groups: group 1 (domain–domain), where the two proteins display well-folded domains, and group 2 (sequence motif-domain), where one of the proteins contains a stretch of residues with a specific binding motif. This information may help in the search for potential target proteins. Dudas et al \cite{dudas2025quantifying} presented a computational protocol for screening molecular libraries to identify new molecular glues based on the application of free energy perturbation to quantify cooperativity. Cooperativity is defined as the ratio between binary and ternary binding affinity, and is expected to be positive for effective molecular glues \cite{wurz2023affinity}.

\begin{figure*}
  \centering
  \includegraphics[width=0.9\textwidth]{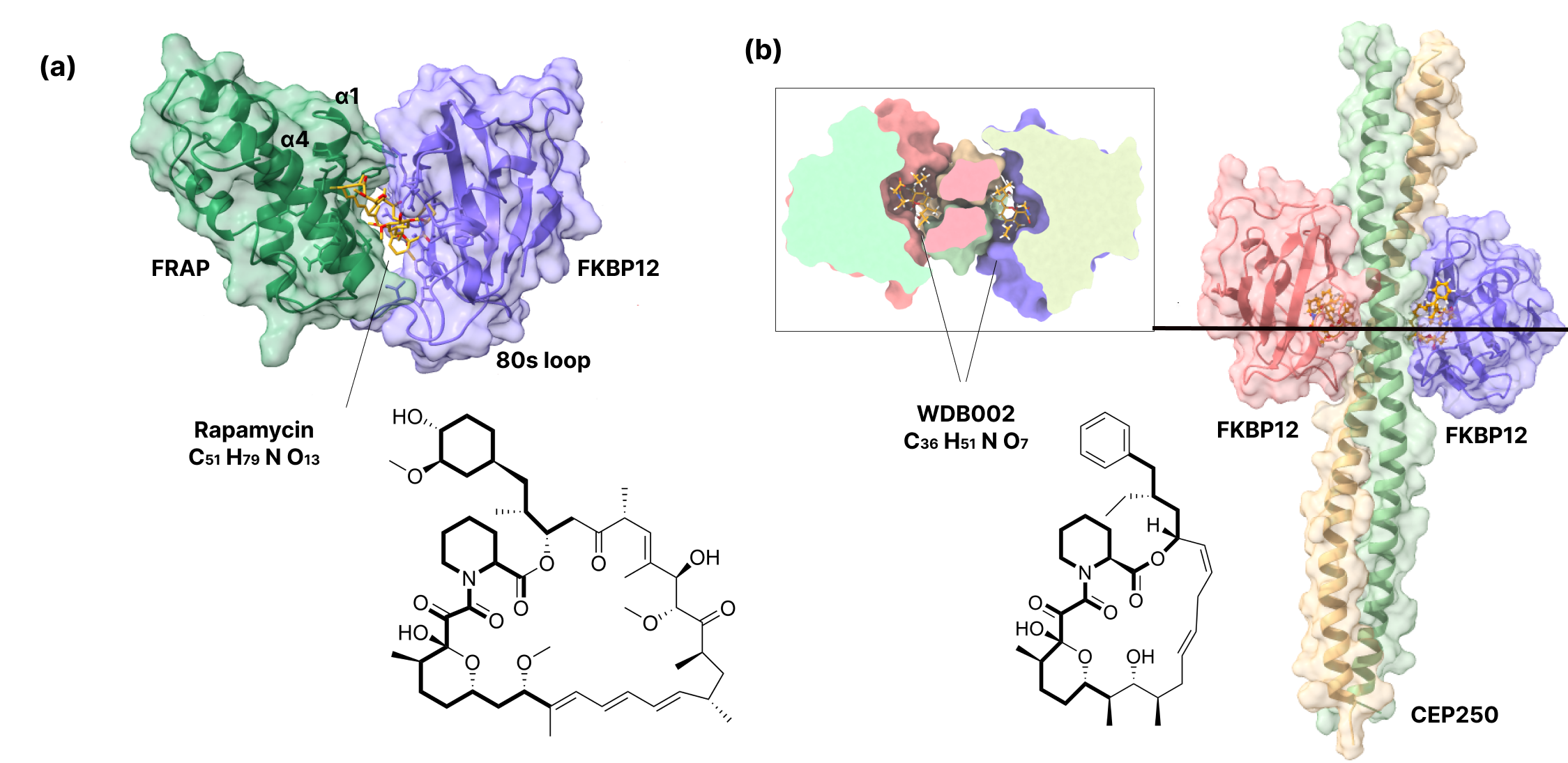}
  \caption{Structures of the ternary complexes (a) FKBP12-FRAP-rapamycin (PDB ID 1FAP \cite{choi1996structure}) and (b) FKBP12-CEP250-WDB002 (PDB ID 6OQA \cite{shigdel2020genomic}), and of the molecular glues rapamycin and WDB002. While the FKBP12-FRAP-rapamycin complex structure has one unit of each molecule (FKBP12, FRAP and rapamycin), the FKBP12-CEP250-WDB002 complex structure contains two units of each molecule (FKBP12, CEP250 and WDB002).}
  \label{fig:struc}
\end{figure*} 

Understanding the factors that contribute to the stability of ternary complexes mediated by molecular glues can lead to mechanistic insights to assist in the rational design of molecular glues for specific effector and target proteins. Several studies employed molecular dynamics (MD) simulations to investigate ternary complexes, providing information about protein interfaces and ligand interactions which assist in the stabilization of ternary complexes \cite{khaledian2025lenalidomide,zhang2024dual}. For instance, Muhammad et al \cite{muhammad2025molecular} recently employed MD simulations to characterize the thermodynamics of ternary complexes involving molecular glues, ubiquitin and the ubiquitin-conjugating enzyme E2, providing a satisfactory ranking of binding affinities and revealing key residues that act as major contributors for molecular glue binding. Chaurasia et al \cite{chaurasia2013molecular} performed MD simulations of the FKBP12-FRB-rapamycin complex to determine the interactions which hold the ternary complex in place. They noted that the molecular glue forms hydrogen bonds with both proteins, but there are only two direct interactions between the proteins, from FKBP12 residues G87 and R43, and the importance of the interacting residues was highlighted with binding free energy calculations. Another study employed MD simulations to investigate the stabilizing forces of ternary complexes, and used this information to devise a virtual screening strategy for potential mTOR inhibitors \cite{kist2018searching}. The focus of these studies has been on the quantification of binding free energies, the characterization of the binding mechanisms and residues of the hotspots on protein interface involved with hydrogen bonding, polar and non-polar interactions. Although the contributions of water at protein interface for ligand or protein binding is well known \cite{zsido2021role, levy2006water, d2024computational}, and the importance of water for mediating protein-protein contacts was established early on \cite{rodier2005hydration}, few studies have explored the role of water and hydration in ternary complexes \cite{chen2025cooperative}. Even though specific water-mediated hydrogen bonds (H-bonds) are present in crystallographic structures for ternary complexes with molecular glues, such as that of FKBP12-rapamycin-FRAP \cite{choi1996structure}, the effects of molecular glues over water-mediated H-bonds have not been explored yet.

In this work, we used MD simulations to investigate two ternary complexes involving molecular glues and FKBP12, FKBP12-rapamycin-FRAP and FKBP12-WDB0002-CEP250 (figure \ref{fig:struc}), and explored how the absence of molecular glues affects protein flexibility and PPIs. The computed binding free energies were in qualitative agreement with previous experiments and showed that the formation of the ternary complexes is thermodynamically favorable. Data analysis of the MD simulations revealed that the molecular glues modulate protein flexibility, leading some protein regions to become less flexible, in agreement with previous experiments. Additionally, the molecular glues changed the pattern and stability of water-mediated H-bonds between target and effector protein, a feature that can be exploited to design optimized molecular glues.


\section{Results and Discussion}

We performed conventional MD simulations of two ternary complexes involving FKBP12, FKBP12-rapamycin-FRAP and FKBP12-WDB0002-CEP250 (figure \ref{fig:struc}), in the presence and absence of the molecular glue (holo and apo systems, respectively) to investigate how the molecular glue affects protein flexibility, complex stability and water-mediated interactions. While the FKBP12-FRAP-rapamycin complex structure has one unit of each molecule (FKBP12, FRAP and rapamycin), the FKBP12-CEP250-WDB002 complex structure contains two units of each molecule (FKBP12, CEP250 and WDB002).
Three replica simulations of the apo and holo systems, and of each of the binary protein-ligand complexes, were performed for 500 ns, as summarized in table \ref{tab:systems}. 

\begin{table}[t]
\centering
\scriptsize
\caption{Summary of systems used to perform molecular dynamics simulations.}
\label{tab:systems}
\begin{tabularx}{8cm}{@{}p{2.5cm}X@{}}
\toprule
{\normalsize \textbf{Ternary }} & {\normalsize \textbf{Systems}} \\
{\normalsize \textbf{Complex}} & \\
\midrule
\textbf{FKBP12–FRAP– }  & FKBP12–FRAP–rapamycin (holo) \\
\textbf{rapamycin}      & FKBP12–FRAP (apo) \\
\textbf{(PDB 1FAP)}         & FKBP12–rapamycin \\
                        & FRAP–rapamycin \\
\addlinespace[3pt]
\textbf{FKBP12–CEP250–} & FKBP12–CEP250–WDB002 (holo)$^a$ \\
\textbf{WDB002}         & FKBP12–CEP250 (apo)$^a$ \\
\textbf{(PDB 6OQA)}         & FKBP12–WDB002$^b$ \\
                        & CEP250–WDB002$^c$ \\
\bottomrule

\end{tabularx}
\begin{tablenotes}
\footnotesize
\item \textit{$^a$}The FKBP12-CEP250-WDB002 complex structure contains two units of each molecule. The two units of each molecule mentioned were included. 
\item \textit{$^b$}Only one unit of FKBP12 (chain A) and one unit of its associated molecular glue (chain A) were included. 
\item \textit{$^c$}Both units of CEP250 (chains C and D) and the molecular glue associated with chain A were included.
\end{tablenotes}
\end{table}

\subsection{Computed binding free energies indicate that the formation of ternary complexes is thermodynamically favorable}

We employed the molecular mechanics / generalized Born surface area (MM/GBSA) method to compute binding free energies for the binary protein-molecular glue complexes and for the ternary complexes (table \ref{tab:mmgbsa}).
The natural molecular glues rapamycin and  WDB002 (figure \ref{fig:struc}), stabilize the PPIs between FKBP12 and the target proteins FRAP and CEP250, respectively \cite{banaszynski2005characterization,shigdel2020genomic}. Previous work used nuclear magnetic resonance (NMR) experiments to investigate the formation of the FKBP12-rapamycin-FRAP complex, showing that rapamycin binds to FKBP12 first, which leads to a conformational change at the protein interaction interface, making it rigid and facilitating the formation of the ternary complex \cite{sapienza2011multi}. Based on this, we computed binding free energies for the formation of the ternary complex using as a ligand the target protein (FRAP or CEP250) and as a receptor the pre-formed FKBP12-molecular glue complex.

The formation of the ternary complex is considered thermodynamically favorable when the binding free energy for the formation of the ternary complex is lower, or more favorable, than the binding free energy for the formation of the binary complexes between the molecular glue and the effector or target protein \cite{li2022importance}.
The binding free energy for the formation of the ternary complex FKBP12-rapamycin-FRAP, of $-61.15 \pm 5.30$ kcal/mol, was lower than that of the associated binary complexes, $-49.48\pm3.95$ and $-34.82 \pm 0.70$ kcal/mol for FKBP12-rapamycin and FRAP-rapamycin, respectively (table \ref{tab:mmgbsa}). Similarly, the binding free energy of the ternary complex FKBP12-WDB0002-CEP250, of $-63.34\pm1.66 $ kcal/mol, was lower than that of the associated binary complexes, $-50.26 \pm2.72$ kcal/mol and $-20.82 \pm0.70$ kcal/mol for FKBP12-WDB0002 and CEP250-WDB0002, respectively (table \ref{tab:mmgbsa}). In both cases, the formation of the ternary complex can be considered thermodynamically favorable, as expected. 

The experimentally determined dissociation constants (\(K_{d}\)) at 25°C for FKBP12-rapamycin-FRAP and FKBP12-WDB002-CEP250 were $12.0 \pm 0.8$ nM \cite{banaszynski2005characterization} and $41.2 \pm 5.7$ nM \cite{shigdel2020genomic}, respectively, resulting in binding free energies of $-10.80 \pm 0.04$ and $-10.07 \pm 0.08$ kcal/mol. The experimental and computed binding free energies for the ternary complexes follow the same qualitative trend, as they have comparable values for the different ternary complexes. We note, however, that the values of experimental and computed binding free energies are different, likely due to approximations involved in the MM/GBSA method \cite{tuccinardi2021current}. 

\begin{table*}
\centering
\caption{Binding free energies ($\Delta G_{bind}$) and cooperative free energies ($\Delta G _{coop}^0 $) computed using the molecular mechanics / generalized Born surface area (MM/GBSA) method  indicate that the formation of the ternary complexes is thermodynamically favorable. Averages and standard deviations from 3 replica simulations of 500 ns for each system.}
\label{tab:mmgbsa}
\begin{tabular}{|lllll|}
\hline
 \begin{tabular}[c]{@{}l@{}}MM/GBSA\\ Receptor \end{tabular} & \begin{tabular}[c]{@{}l@{}}MM/GBSA \\ Ligand \end{tabular}  & \begin{tabular}[c]{@{}l@{}}Binary Complex \\ $\Delta G_{bind}$  (kcal/mol) \end{tabular} & 
 \begin{tabular}[c]{@{}l@{}}Ternary Complex \\ $\Delta G_{bind}$ (kcal/mol)\end{tabular} & \begin{tabular}[c]{@{}l@{}} Ternary Complex \\ $\Delta G _{coop}^0 $ (kcal/mol)\end{tabular}\\
\hline
FKBP12 & Rapamycin      & $-49.48 \pm  3.95$ &  &             \\
FRAP & Rapamycin        & $-34.82 \pm 0.70 $  &  &             \\
FKBP12-Rapamycin & FRAP &  & $-61.15  \pm 5.30$ & $ -26.33 \pm 5.35$ \\
\hline
FKBP12 & WDB002         & $-50.26 \pm 2.72$    & &              \\
CEP250 & WDB002         & $-20.82 \pm 0.70 $   &   &            \\
FKBP12-WDB002 & CEP250 &                  & $-63.34 \pm 1.66$ & $ -42.52 \pm 1.80 $ \\
\hline
\end{tabular}
\end{table*}

In the investigation of the stability of ternary complexes, it is common to not only quantify binding free energies, but also
 cooperativity  \cite{dudas2025quantifying,chen2025cooperative}. Following previous work \cite{dudas2025quantifying}, the cooperative free energy ($\Delta G _{coop}^0$) is defined as the difference between the binding free energies of the ternary complex and of the binary complex target protein-molecular glue. From our computed binding free energies we observe favorable (negative) $\Delta G _{coop}^0$ values for both ternary complexes (table \ref{tab:mmgbsa}), indicating positive cooperativity, as expected from the stably bound ternary complexes studied here.   

\subsection {The molecular glue modulates protein flexibility}

When rapamycin binds to the cavity of FKBP12, it inhibits the rotamase activity of the enzyme \cite{harding1989receptor}. According to previous NMR experiments, rapamycin binding turns the backbone of FKBP12 more rigid in the regions in contact with the molecular glue, namely the 80's loop along with residues E55, G59, F37 and D38\cite{sapienza2011multi}. MD simulations of the ternary complexes in apo and holo conditions were conducted to further investigate the effect of the molecular glue over the protein conformations. 

\begin{figure*}
  \centering
  \includegraphics[width=17.4cm]{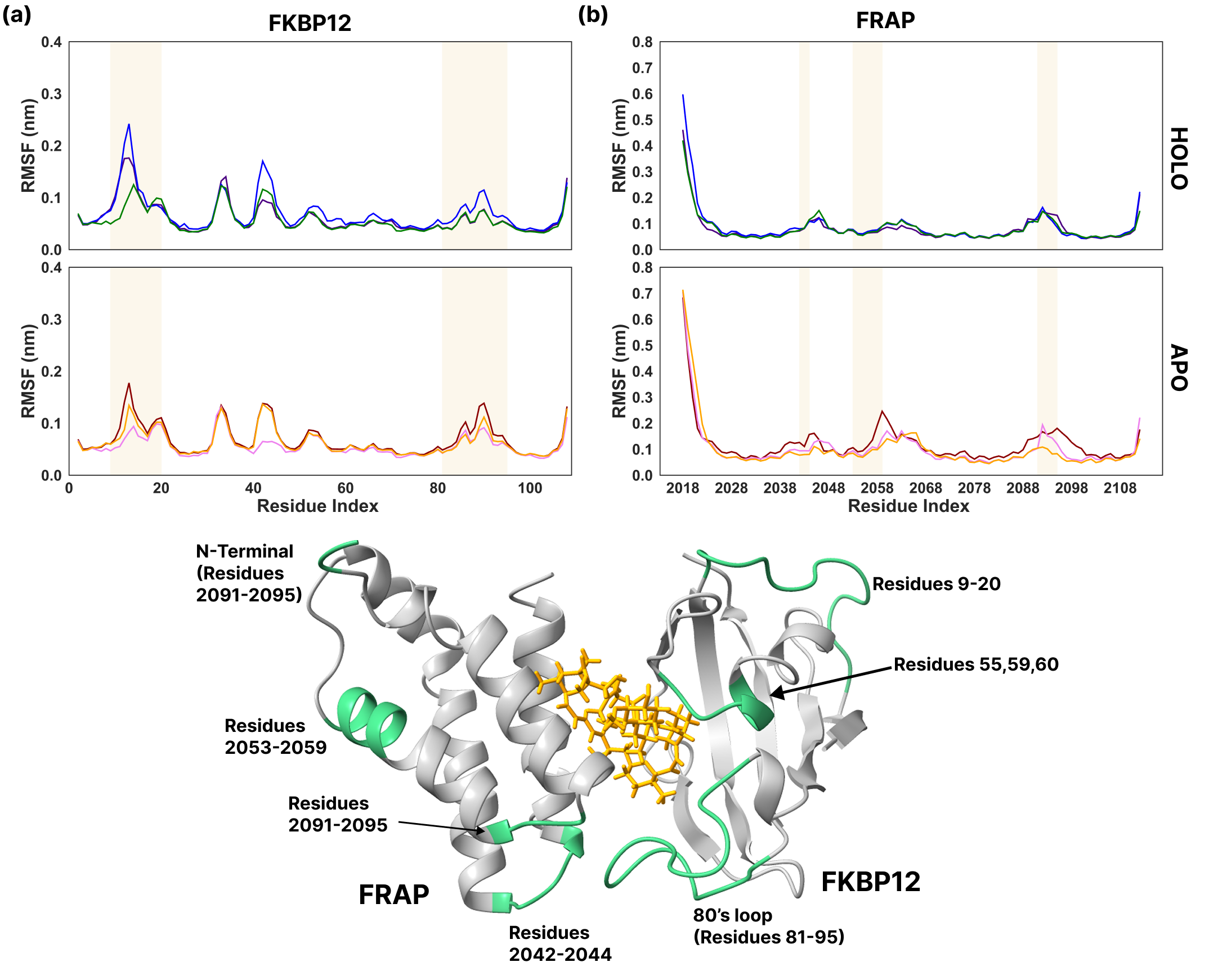}
  \caption{Root mean square fluctuation (RMSF) of backbone atoms, after structure alignment with the backbone of FKBP12 (a) and FRAP (b) in  holo (top row) and apo (bottom row) conditions from three replica MD simulations of 500 ns of the FKBP12-FRAP-rapamycin complex. In the holo condition, the 80s loop of FKBP12 becomes less flexible. The loop regions of FRAP at the protein-protein interaction interface become more flexible in the apo condition. Shaded regions indicate differences in the RMSF values in apo and holo conditions. Bottom: structure of the FKBP12-FRAP-rapamycin complex (PDB 1FAP), with regions with modified RMSF values highlighted in green. }
  \label{fig:rmsf-frap}
\end{figure*}

\begin{figure*}
  \centering
  \includegraphics[width=0.9\textwidth]{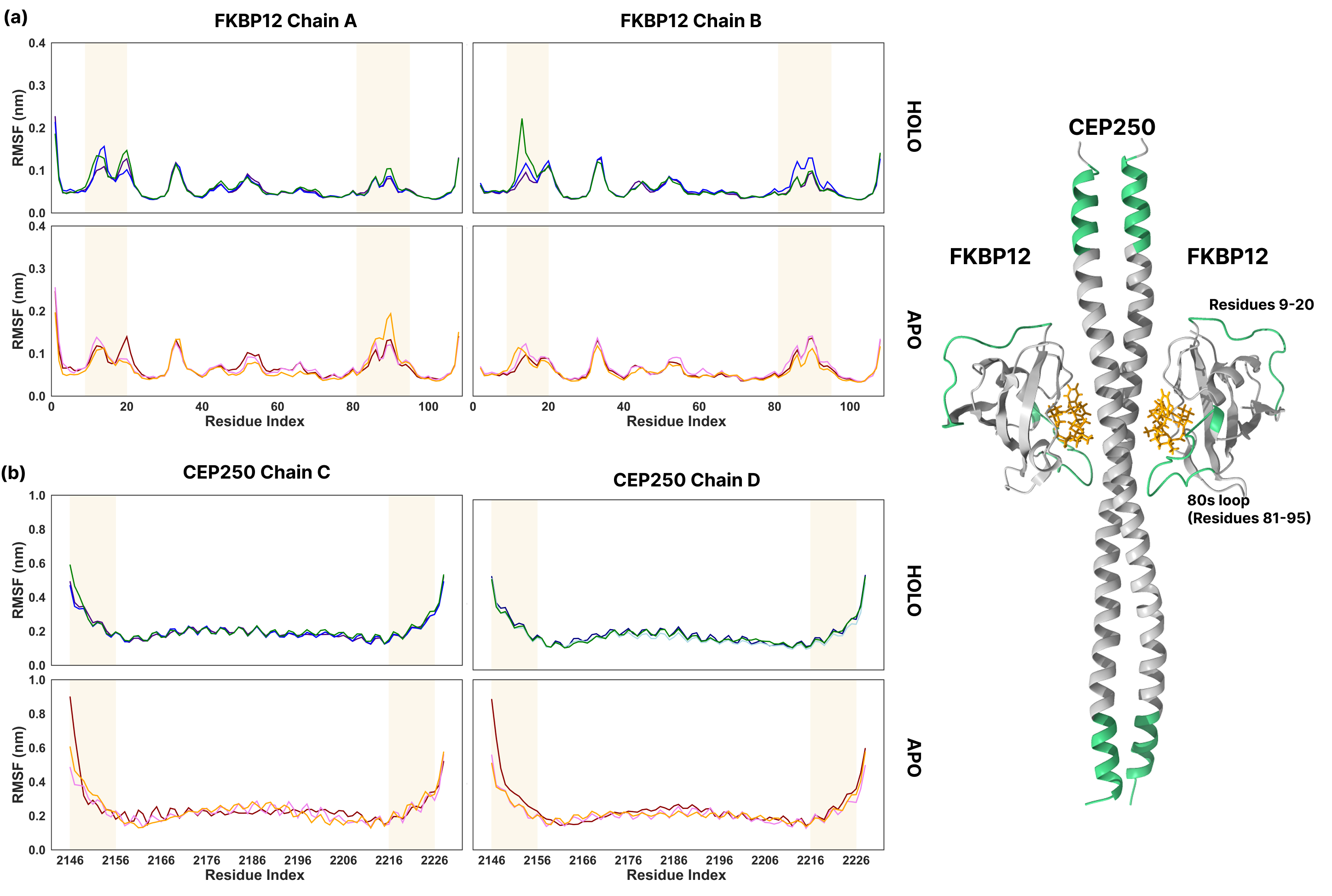}
  \caption{Root mean square fluctuation (RMSF) of backbone atoms, after structure alignment with the backbone of FKBP12 (a) and CEP250 (b) in holo (top row) and apo (bottom row) conditions from three replica MD simulations of 500 ns of the FKBP12-CEP250-WDB002 complex. In the holo condition, the 80s loop of FKBP12 becomes less flexible. The terminal residues in CEP250 fluctuate more in the apo condition. Shaded regions indicate differences in the RMSF values in apo and holo conditions. Right: structure of the FKBP12-CEP250-WDB002 complex (PDB 6OQA), with regions with modified RMSF values highlighted in green.}
  \label{fig:rmsf-cep}
\end{figure*}

In the FKBP12-FRAP-rapamycin simulations we observed that the 80s loop backbone (residues 81-95) of FKBP12 becomes less flexible (lower root mean squared fluctuation, RMSF) in the holo condition (figure \ref{fig:rmsf-frap}), in agreement with previous NMR experiments \cite{sapienza2011multi}. Additionally, we see the outer loop (residues 9-20) is more flexible in holo conditions in at least two replicas of the holo simulations (figure \ref{fig:rmsf-frap}). These conformational changes in FKBP12 are also visible in the apo and holo conditions of the FKBP12-CEP250-WDB002 ternary complex (figure \ref{fig:rmsf-cep}). No clear differences were observed between the apo and holo conditions in the root mean square deviations (RMSD) for FKBP12 (Figure S1). 

Lower flexibility of the residues of the target proteins, FRAP and CEP250, was also observed in holo simulations. FRAP has two loops (residues 2042-2044 and 2091-2095), located between the \(\alpha1\) and \(\alpha4\) helices and at the protein-protein interaction interface, and both loops were less flexible in the holo condition (figure \ref{fig:rmsf-frap}). In addition, the N-terminal and the distal alpha helix (residues 2053-2059) were less flexible in the holo condition. In the case of CEP250, there are no loops and the residues at the interaction interface did not show clear differences in flexibility in the apo and holo conditions. Higher flexibility of the N and C-termini was observed in the apo condition (figure \ref{fig:rmsf-cep}). This change results from the unwinding of the coiled CEP250 dimer in the absence of the molecular glue. No clear differences were observed between the apo and holo conditions in the RMSD values for the target proteins (Figure S2). 

When the average minimum distance between atoms of FKBP12 and the target protein were measured in holo and apo conditions, we observed that the proteins moved closer by a few Angstroms when the molecular glue was removed from the FKBP12-FRAP-rapamycin complex (Figure S5), indicating a structural role of the molecular glue, which helps to keep the proteins in place in the ternary complex. This is a consequence of the proteins not displaying major direct interactions in this complex \cite{choi1996structure}.  Since the FKBP12-CEP250-WDB002 complex is also held together by direct H-bonds between the proteins \cite{shigdel2020genomic}, the absence of a molecular glue did not cause major changes in the inter-protein distance, but led to larger standard deviations (Figure S5). 

In agreement with previous experimental work \cite{sapienza2011multi}, we can observe that the binding of the molecular glue modulates protein flexilibility, turning some regions more rigid, which may facilitate the formation of an interface for protein-protein interactions.
Modifications of such interfaces have also been observed in other ternary complexes, as reviewed by Robinson et al \cite{robinson2024molecular}.

\subsection{The molecular glue stabilizes water-mediated hydrogen bonds between proteins}

Understanding the nature of the protein-protein interaction interfaces provides insights into the determinants of complex stability. The FKBP12-molecular glue and target protein interface is mainly composed of hydrophobic residues \cite{shigdel2020genomic}. FKBP12 binds to molecular glues and target proteins through specific conserved residues (K45, F47, K48, A82, Y83, T86, G87, H88, P89, G90, I91). Most of these residues are located in the 80s loop, which engages in direct H-bond interactions with the target protein after undergoing target-mediated conformational changes \cite{shigdel2020genomic}. First, we investigated the number of direct H-bonds between proteins in apo and holo conditions. While direct H-bonds in the complex are usually considered as important for strong binding affinities and stability, we observed no major differences in the number of direct H-bonds in the apo and holo complexes (Figures S3 and S4).

\begin{figure*}
  \centering
  \includegraphics[width=17.4cm]{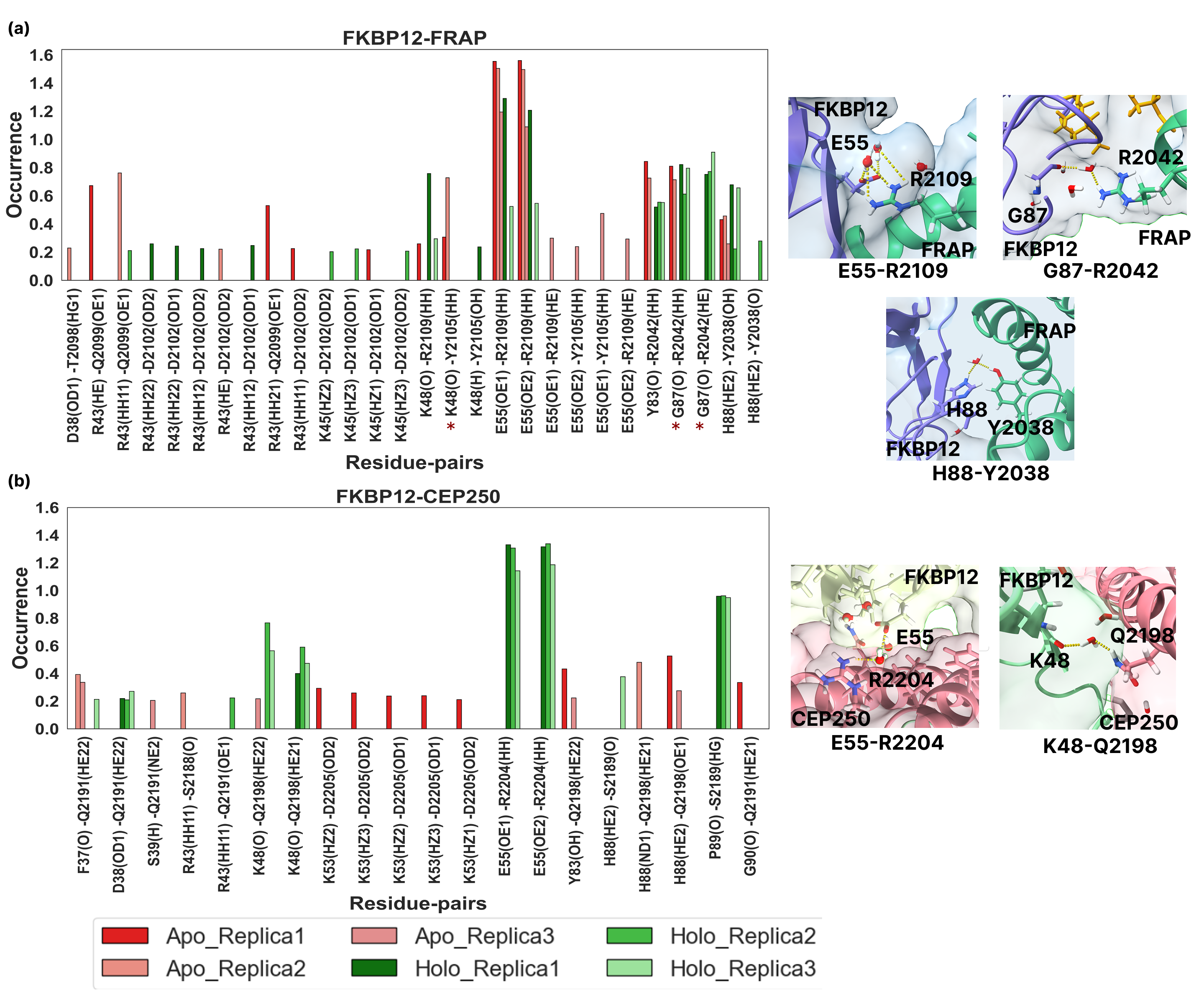}
  \caption{Frequency of occurrence of water-mediated hydrogen bonds (H-bonds) between (a) FKBP12 and FRAP of the FKBP12-FRAP-rapamycin complex and (b) FKBP12 and CEP250 of the FKBP12-WDB002-CEP250 complex in apo and holo conditions, obtained from three replica MD simulations of 500 ns. Naming scheme: first atom in FBKP12, second atom in the target protein (FRAP or CEP250). The water-mediated H-bonds Lys48-Tyr2105, Gly87-Arg2042 and Thr89-Arg2042 (which does not appear for more than 20\% of simulation length) were also present in crystallographic structure of the FKBP12-FRAP-rapamycin complex. The water-mediated H-bonds were identified using MDAnalysis with 3 Å as distance cut-off (between donor and acceptor atoms) and 120º as angle cut-off, and a frequency higher than 1 can happen when two water molecules interact with the same atom pair. Only interactions which lasted for at least 20\% of the simulation length were plotted. Right: water-mediated H-bonds from representative snapshots obtained from MD simulations. Yellow lines indicate H-bonds.}
  \label{fig:heatmap-frap}
\end{figure*}

Next, motivated by the presence of three water-mediated H-bonds (Lys48-Tyr2105, Thr89-Arg2042 and Gly87-Arg2042, first residue in FKBP12, second residue in FRAP) in the crystallographic structure of the FKBP12-FRAP-rapamycin complex, we investigated the number and stability of water-mediated H-bonds between proteins in apo and holo conditions. Water-mediated H-bonds of order 2 ,i.e., H-bonds mediated by a maximum of 2 water molecules \cite{garcia2013hydration}, were considered, and their frequency of occurrence in each simulation was measured. A frequency of occurrence over 1 indicates that two separate water molecules formed a water-mediated H-bond with the same set of atoms. Only interactions which lasted for at least 20\% of the simulation length are shown. The differences in the number of water-mediated H-bonds and their frequency of occurrence were noticeable. When comparing the apo and holo conditions, it is clear from figure \ref{fig:heatmap-frap} that the water-mediated H-bonds are different in apo and holo conditions.
In the FKBP12-rapamycin-FRAP complex, two out of three water-mediated H-bonds present in the crystallographic structure had a higher frequency of occurrence, being more stable in the holo condition (figure \ref{fig:heatmap-frap}a).
In the FKBP12-CEP250-WDB002 complex, more water-mediated H-bonds are present in the holo condition, and these H-bonds also have a higher frequency occurrence, or more stability, in the MD simulations in the holo condition (figure \ref{fig:heatmap-frap}b). Binding of the molecule glue reduces the flexibility of the 80s loop in FKBP12, facilitating the formation of water-mediated H-bonds with residues in this region, such as Gly87 and Pro89. Additionally, it may be possible that the molecular glue facilitates and stabilizes water-mediated interactions by bringing the two proteins at a specific distance from each other (figure S5).

The residues of the 40s (residues 38-48) and 80s loop of FKBP12 were seen to participate in water-mediated H-bonds when the molecular glue is present. As seen in figure \ref{fig:heatmap-frap}, three residues of FKBP12 were found to be involved in water-mediated H-bonds in both the FKBP12-rapamycin-FRAP and FKBP12-WDB002-CEP250 ternary complexes, Arg42, Lys47 and Glu54. Meanwhile, another 4 or 5 residues from FKBP12 were involved in water-mediated H-bonds in only one of the ternary complexes.

As observed from crystal structures \cite{rodier2005hydration}, water often permeates protein-protein interfaces and mediates protein-protein interactions, and the hydration shell of a protein is known to play an important role in protein function and molecule recognition \cite{fogarty2014water}. This could be observed in our MD simulations as well, where certain water-mediated interactions were more stable in the presence of the molecular glues, as in the case of FKBP12-FRAP-rapamycin, and even increase in number, as in the case of FKBP12-CEP250-WDB002.

\section{Conclusion}

Molecular glues hold the promise of a new category of therapeutic agents, but studies are still required for a complete understanding of how they stabilize PPIs across different ternary complexes. Here, we used MD simulations to provide mechanistic insights into the stabilizing effects of the molecular glues rapamycin and WDB002 over the ternary complexes formed with FKBP12 and FRAP or CEP250, respectively. We computed binding free energies for the ternary complexes and related binary complexes, showing that the formation of the ternary complex is thermodynamically favorable, and that the computed results have qualitative agreement with previous experiments. Additional analysis revealed consistent behavior for the different ternary complexes. We observed that the presence of the molecular glue changes protein flexibility, usually leading to reduced flexibility in specific regions of the proteins involved in the ternary complex, in agreement with experimental results previously published. Interestingly, we also observed that the water-mediated H-bonds are very sensitive to the presence of the molecular glue. Our results show that the pattern and stability of water-mediated H-bonds is modified in the presence of molecular glues, which contribute to turn some protein regions in FKBP12 (particularly the 80s loop) more rigid, therefore facilitating the formation of such H-bonds. 

While there are many computational methods that investigate solvation and water-mediated interactions in binary complexes, future work could investigate whether such methods can also be applied for ternary complexes. Additionally, our results indicate that water-mediated H-bonds may play a role in ternary complex stabilization, and should be considered in efforts to model ternary complexes, and in the design and optimization of molecular glues.  

\section*{Supporting Information}

Description of the computational methods used in this work and additional results (protein root mean square deviation, number of hydrogen bonds between proteins, and pairwise distance between proteins in different MD simulations) can be found within the Supporting Information.

\section*{Acknowledgements}

A.M. and A.N.A. thank funding from DFG under Germany’s Excellence Strategy – EXC 2008/1-390540038 – UniSysCat.
M.A.C. thanks funding from Erasmus+ grant and from Universitat Politècnica de Catalunya, Spain.

\section*{Conflict of Interest}

The authors declare no conflict of interest.

\begin{shaded}
\noindent\textsf{\textbf{Keywords:} \keywords} 
\end{shaded}

\setlength{\bibsep}{0.0cm}
\bibliographystyle{Wiley-chemistry}
\bibliography{manuscript}

\begin{thebibliography}{10}

\bibitem{konstantinidou2024molecular}
M.~Konstantinidou, M.~R. Arkin, \emph{Cell Chemical Biology} \textbf{2024}, \emph{31}, 1064.

\bibitem{nada2024new}
H.~Nada, Y.~Choi, S.~Kim, K.~S. Jeong, N.~A. Meanwell, K.~Lee, \emph{Signal Transduction and Targeted Therapy} \textbf{2024}, \emph{9}, 341.

\bibitem{gao2020protac}
H.~Gao, X.~Sun, Y.~Rao, \emph{ACS medicinal chemistry letters} \textbf{2020}, \emph{11}, 237.

\bibitem{as2024protac}
B.~G. AS, D.~Agrawal, N.~M. Kulkarni, R.~Vetrivel, K.~Gurram, \emph{ACS omega} \textbf{2024}, \emph{9}, 12611.

\bibitem{sun2019protacs}
X.~Sun, H.~Gao, Y.~Yang, M.~He, Y.~Wu, Y.~Song, Y.~Tong, Y.~Rao, \emph{Signal transduction and targeted therapy} \textbf{2019}, \emph{4}, 64.

\bibitem{tomlinson2025three}
A.~C. Tomlinson, J.~E. Knox, L.~Brunsveld, C.~Ottmann, J.~K. Yano, \emph{Current Opinion in Structural Biology} \textbf{2025}, \emph{91}, 103007.

\bibitem{holdgate2024screening}
G.~A. Holdgate, C.~Bardelle, S.~K. Berry, A.~Lanne, M.~E. Cuomo, \emph{SLAS Discovery} \textbf{2024}, \emph{29}, 100136.

\bibitem{xie2023recent}
X.~Xie, T.~Yu, X.~Li, N.~Zhang, L.~J. Foster, C.~Peng, W.~Huang, G.~He, \emph{Signal transduction and targeted therapy} \textbf{2023}, \emph{8}, 335.

\bibitem{banaszynski2005characterization}
L.~A. Banaszynski, C.~W. Liu, T.~J. Wandless, \emph{Journal of the American Chemical Society} \textbf{2005}, \emph{127}, 4715.

\bibitem{siekierka1989cytosolic}
J.~J. Siekierka, S.~H. Hung, M.~Poe, C.~S. Lin, N.~H. Sigal, \emph{Nature} \textbf{1989}, \emph{341}, 755.

\bibitem{choi1996structure}
J.~Choi, J.~Chen, S.~L. Schreiber, J.~Clardy, \emph{Science} \textbf{1996}, \emph{273}, 239.

\bibitem{shigdel2020genomic}
U.~K. Shigdel, S.-J. Lee, M.~E. Sowa, B.~R. Bowman, K.~Robison, M.~Zhou, K.~H. Pua, D.~T. Stiles, J.~A. Blodgett, D.~W. Udwary, et~al., \emph{Proceedings of the National Academy of Sciences} \textbf{2020}, \emph{117}, 17195.

\bibitem{nunes2023alphafold2}
A.~Nunes-Alves, K.~Merz, \emph{Journal of Chemical Information and Modeling} \textbf{2023}, \emph{63}, 5947.

\bibitem{salo2020molecular}
O.~M. Salo-Ahen, I.~Alanko, R.~Bhadane, A.~M. Bonvin, R.~V. Honorato, S.~Hossain, A.~H. Juffer, A.~Kabedev, M.~Lahtela-Kakkonen, A.~S. Larsen, et~al., \emph{Processes} \textbf{2020}, \emph{9}, 71.

\bibitem{schimunek2024community}
J.~Schimunek, P.~Seidl, K.~Elez, T.~Hempel, T.~Le, F.~No{\'e}, S.~Olsson, L.~Raich, R.~Winter, H.~Gokcan, et~al., \emph{Molecular informatics} \textbf{2024}, \emph{43}, e202300262.

\bibitem{rui2023protein}
H.~Rui, K.~S. Ashton, J.~Min, C.~Wang, P.~R. Potts, \emph{RSC Chemical Biology} \textbf{2023}, \emph{4}, 192.

\bibitem{ben2025molecular}
A.~Ben~Geoffrey, D.~Agrawal, N.~M. Kulkarni, M.~Gunasekaran, \emph{ACS omega} \textbf{2025}, \emph{10}, 6650.

\bibitem{xu2025accommodating}
Q.~Xu, W.~Liu, H.~Liu, Z.~Zheng, \emph{Journal of Chemical Information and Modeling} \textbf{2025}.

\bibitem{deutscher2025discovery}
R.~C. Deutscher, C.~Meyners, M.~L. Repity, W.~O. Sugiarto, J.~M. Kolos, E.~V. Maciel, T.~Heymann, T.~M. Geiger, S.~Knapp, F.~Lermyte, et~al., \emph{Chemical science} \textbf{2025}, \emph{16}, 4256.

\bibitem{lukauskis2025optimizing}
D.~Lukauskis, N.~Kashif-Khan, C.~Tame, A.~Potterton \textbf{2025}.

\bibitem{dudas2025quantifying}
B.~Dudas, C.~Athanasiou, J.~C. Mobarec, E.~Rosta, \emph{Journal of Chemical Theory and Computation} \textbf{2025}, \emph{21}, 5712.

\bibitem{wurz2023affinity}
R.~P. Wurz, H.~Rui, K.~Dellamaggiore, S.~Ghimire-Rijal, K.~Choi, K.~Smither, A.~Amegadzie, N.~Chen, X.~Li, A.~Banerjee, et~al., \emph{Nature communications} \textbf{2023}, \emph{14}, 4177.

\bibitem{khaledian2025lenalidomide}
M.~Khaledian, A.~Divsalar, F.~Badalkhani-Khamseh, A.~A. Saboury, B.~Ghalandari, X.~Ding, M.~Zamanian-Azodi, \emph{Journal of Molecular Liquids} \textbf{2025}, \emph{431}, 127717.

\bibitem{zhang2024dual}
Z.~Zhang, Y.~Li, J.~Yang, J.~Li, X.~Lin, T.~Liu, S.~Yang, J.~Lin, S.~Xue, J.~Yu, et~al., \emph{Nature Communications} \textbf{2024}, \emph{15}, 6477.

\bibitem{muhammad2025molecular}
D.~Muhammad, W.~Xia, M.~Wang, Z.~Sun, J.~Z. Zhang, \emph{International Journal of Biological Macromolecules} \textbf{2025}, \emph{306}, 141454.

\bibitem{chaurasia2013molecular}
S.~Chaurasia, S.~Pieraccini, R.~De~Gonda, S.~Conti, M.~Sironi, \emph{Chemical Physics Letters} \textbf{2013}, \emph{587}, 68.

\bibitem{kist2018searching}
R.~Kist, L.~F. S.~M. Timmers, R.~A. Caceres, \emph{Journal of Molecular Graphics and Modelling} \textbf{2018}, \emph{80}, 251.

\bibitem{zsido2021role}
B.~Z. Zsid{\'o}, C.~Het{\'e}nyi, \emph{Current Opinion in Structural Biology} \textbf{2021}, \emph{67}, 1.

\bibitem{levy2006water}
Y.~Levy, J.~N. Onuchic, \emph{Annu. Rev. Biophys. Biomol. Struct.} \textbf{2006}, \emph{35}, 389.

\bibitem{d2024computational}
G.~D’Arrigo, D.~B. Kokh, A.~Nunes-Alves, R.~C. Wade, \emph{Communications Biology} \textbf{2024}, \emph{7}, 1159.

\bibitem{rodier2005hydration}
F.~Rodier, R.~P. Bahadur, P.~Chakrabarti, J.~Janin, \emph{Proteins: Structure, Function, and Bioinformatics} \textbf{2005}, \emph{60}, 36.

\bibitem{chen2025cooperative}
S.-Y. Chen, R.~Solazzo, M.~Fouch{\'e}, H.-J. Roth, B.~Dittrich, S.~Riniker, \emph{Journal of chemical theory and computation} \textbf{2025}, \emph{21}, 8557.

\bibitem{sapienza2011multi}
P.~J. Sapienza, R.~V. Mauldin, A.~L. Lee, \emph{Journal of molecular biology} \textbf{2011}, \emph{405}, 378.

\bibitem{li2022importance}
W.~Li, J.~Zhang, L.~Guo, Q.~Wang, \emph{Journal of Chemical Information and Modeling} \textbf{2022}, \emph{62}, 523.

\bibitem{tuccinardi2021current}
T.~Tuccinardi, \emph{Expert opinion on drug discovery} \textbf{2021}, \emph{16}, 1233.

\bibitem{harding1989receptor}
M.~W. Harding, A.~Galat, D.~E. Uehling, S.~L. Schreiber, \emph{Nature} \textbf{1989}, \emph{341}, 758.

\bibitem{robinson2024molecular}
S.~A. Robinson, S.~M. Banik, et~al., \emph{Cell Chemical Biology} \textbf{2024}, \emph{31}, 1089.

\bibitem{garcia2013hydration}
A.~T. Garc{\'\i}a-Sosa, \emph{Journal of chemical information and modeling} \textbf{2013}, \emph{53}, 1388.

\bibitem{fogarty2014water}
A.~C. Fogarty, D.~Laage, \emph{The Journal of Physical Chemistry B} \textbf{2014}, \emph{118}, 7715.

\end{thebibliography}









\end{document}